\documentclass[twocolumn,aps,prl,superscriptaddress,longbibliography]{revtex4-2}

\pdfoutput=1

\usepackage{amssymb}
\usepackage{amsmath}
\usepackage{graphicx,bm}
\usepackage{epstopdf}
\usepackage{subfigure}
\usepackage{natbib}
\usepackage{epsfig}
\usepackage{amsfonts}
\usepackage{mathrsfs}
\usepackage{CJK}
\usepackage[toc,page,title,titletoc,header]{appendix}
\usepackage{array,longtable}
\usepackage{float}
\usepackage[usenames, dvipsnames, x11names]{xcolor}
\usepackage[colorlinks]{hyperref}
\usepackage[capitalise]{cleveref}
\usepackage{soul}
\hypersetup{
 colorlinks=true,
 citecolor=red,
 linkcolor=blue,
 urlcolor=cyan}
\usepackage{mathtools}

\newcommand{\sgn}{ \ \mathrm{sgn}}

\renewcommand{\H}{\hat{\mathcal{H}}}
\newcommand{\D}{\mathcal{D}}

\renewcommand{\c}[1]{\hat{c}_{\overline{#1}}}
\newcommand{\0}{\overline{0}}
\newcommand{\1}{\overline{1}}

\newcommand{\betaA}{\hat{\beta}_L}

\DeclareMathOperator{\hc}{H.c.}

\usepackage{comment}

\begin{document}
\title{Stabilizing volume-law entangled states of fermions and qubits using local dissipation}

\author{Andrew Pocklington}
\affiliation{Pritzker School of Molecular Engineering,  University  of  Chicago, 5640  South  Ellis  Avenue,  Chicago,  Illinois  60637,  U.S.A.}

\author{Yu-Xin Wang}
\affiliation{Pritzker School of Molecular Engineering,  University  of  Chicago, 5640  South  Ellis  Avenue,  Chicago,  Illinois  60637,  U.S.A.}

\author{Yariv Yanay}
\affiliation{Laboratory for Physical Sciences, 8050 Greenmead Dr., College Park, MD 20740}

\author{A. A. Clerk}
\affiliation{Pritzker School of Molecular Engineering,  University  of  Chicago, 5640  South  Ellis  Avenue,  Chicago,  Illinois  60637,  U.S.A.}

\date{\today}

\begin{abstract}
We analyze a general method for the dissipative preparation and stabilization of volume-law entangled states of fermionic and qubit lattice systems in 1D (and higher dimensions for fermions).  Our approach requires minimal resources:  nearest-neighbour Hamiltonian interactions that obey a suitable chiral symmetry, and the realization of just a single, spatially-localized dissipative pairing interaction.  In the case of a qubit array, the dissipative model we study maps to an interacting fermionic problem.  Nonetheless, we analytically show the existence of a unique pure entangled steady state (a so-called rainbow state).  Our ideas are compatible with a number of experimental platforms, including superconducting circuits and trapped ions.
\end{abstract}

\maketitle

{\textit{Introduction-- }}
Quantum reservoir engineering is a powerful tool in quantum information processing.  In its simplest form, it involves tailoring dissipative processes to stabilize non-classical quantum states \cite{Poyatos1996, Plenio2002}; when generalized to stabilizing a subspace, it can also be used as a route to quantum error correction \cite{Reiter2017,Jiang2021}.  Many experiments have implemented dissipation engineering in few-body quantum systems comprised of 1-2 qubits or bosonic modes 
(see e.g.~\cite{Lin2013,Shankar2013,Wollman2015,Kienzler2015,Kimchi-Schwartz2016}).
Theoretical work has also considered extensions to truly many-body systems \cite{Diehl2008,Kraus2008,Diehl2011}, though most proposals are experimentally daunting, as they require engineered dissipation on every site of an extended lattice system.  More recent work demonstrated that for non-interacting bosons hopping on a 1D lattice, a single, local engineered squeezing dissipator can be sufficient to stabilize the entire extended system in a state with long-range entanglement \cite{Zippilli2015,Ma2017}; a subtle particle-hole symmetry was shown to be the key ingredient, allowing a generalization to higher dimensions \cite{Yanay2018}.  These protocols are, however, limited to stabilizing Gaussian entangled states, whose use in quantum information is highly constrained \cite{Bartlett2002}.

\begin{figure}[t!]
\centering
\includegraphics[width = 3.25in]{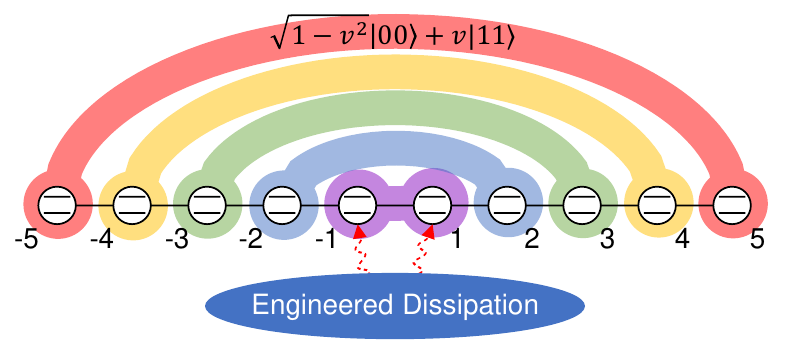}
\caption{
Schematic of a 1D qubit array with nearest-neighbour XY interactions, where the two central sites are coupled to a common engineered dissipative reservoir with a pairing parameter $v$ (c.f.~\cref{eqn:jump}).  The dissipative dynamics stabilizes an arbitrary initial state of the qubits into a volume-law entangled rainbow state, where each qubit is entangled with its mirror image qubit (as depicted).  The model maps to an interacting fermionic model featuring dissipative pairing with phase fluctuations.  
}
\label{fig:rainbow}
\end{figure}

Given this prior work, a natural question is whether a single localized dissipative process can prepare and stabilize more complex many-body entangled states.  In particular, can this approach work in systems which have (unlike free bosons) a finite-dimensional local Hilbert space, e.g.~lattices of fermions, hard-core bosons or qubits. In this Letter, we show that the answer is, surprisingly, yes.  We describe an extremely simple protocol exploiting symmetry and the  dissipative analog of Cooper pairing to stabilize highly entangled states in 1D lattices of fermions and qubits, one example being the so-called ``rainbow state'' (\cref{fig:rainbow}).  Such rainbow states feature long-range, volume-law entanglement, and are known to be the ground states of highly structured, spatially non-uniform Hamiltonians \cite{Vitagliano2010,Zhang2017}.  Our dissipative approach does not require the realization of such exotic Hamiltonians.  Instead, it only uses nearest-neighbour Hamiltonian interactions (which need not be uniform or symmetric) and a {\it single} localized dissipator; the entangled steady state state is the unique steady state irrespective of the size of the lattice. As we discuss, the resources required to implement our protocol already exist in a number of different quantum information processing architectures.  


Our results also have interest in the context of general studies of many-body driven dissipative system. The spin version of our problem {\it cannot be mapped exactly to free fermions}.  Nonetheless, we are able to exactly describe the steady state.  We discuss how qualitative features of the dynamics can be connected to a model of dissipative fermionic pairing with phase fluctuations.  Note that our work is distinct from a recent proposal for using dissipation to generate entangled states in 1D qubit chains \cite{Dutta2020, Dutta2021}.  These protocols also generated rainbow-like entangled states, but only if the system was initially prepared in a non-trivial, highly nonlocal entangled steady state. In contrast, our approach has in general a {\it unique} entangled steady state, and hence is completely independent of the initial state: one can start from a trivial product state and still obtain the volume-law entangled rainbow state.  Alternate schemes that unconditionally stabilize qubit rainbow states have also been proposed \cite{WendenbaumPRA2015, ZipilliPRL2013}.  Our scheme is simpler to implement, and is also far more general: it can stabilize a wide class of entangled pure qubit states (many having a correlation structure  considerably more complex than a rainbow).


{\textit{Fermions-- }}
We begin by considering non-interacting fermions, using this system to build up the key ideas that will enable our qubit protocol.  
We consider spinless fermions hopping on a $2N$-site lattice with a tight binding Hamiltonian $\H_F = \sum_{i,j} H_{ij} \hat{c}_i^\dagger \hat{c}_j$, where $H_{ij}$ is a Hermitian matrix, and $\hat{c}_i$ annihilates a fermion at lattice site $i$.
$\H_F$ is readily diagonalized, with $\hat{d}_\alpha^\dagger = \sum_j \psi_\alpha[j] \hat{c}_j^\dagger$ creating a particle in an energy eigenstate with energy $\epsilon_\alpha$ and real-space wavefunction  $\psi_\alpha[j]$.  Note that we do not assume translational invariance.

Our goal is to now introduce localized dissipation which stabilizes the entire lattice in a finite-density state with long-range entanglement.  For non-interacting bosons, this can be accomplished by coupling a single site to a squeezed Markovian reservoir \cite{Zippilli2015,Ma2017,Yanay2018}.  Such a reservoir attempts to enforce local pairing correlations on the coupled site. For spinless fermions, the Pauli exclusion principle excludes an identical approach.  However, one can try the next simplest configuration: introduce a localized Markovian dissipative reservoir that attempts to stabilize fermionic pairing correlations on {\it two} adjacent sites $j=\0,\1$ (i.e.~prepare them in the state $(u + v e^{i \phi}   \hat{c}^\dagger_{\0} \hat{c}^\dagger_{\1} ) |00 \rangle $). This corresponds to simply cooling a pair of localized fermionic Bogoliubov modes.  As we will see, simply cooling one of these modes generally suffices.  

The total system dynamics including the localized dissipative pairing is then described by a Lindblad master equation:
\begin{align}
    &\dot{\hat{\rho}} = -i[\H,\hat{\rho}] + \Gamma \D[\betaA] \hat{\rho}, \label{eqn:master}
\end{align}
Here $\D[\hat{L}]\hat{\rho} = \hat{L} \hat{\rho} \hat{L}^\dagger - \frac{1}{2} \{\hat{L}^\dagger \hat{L}, \hat{\rho} \}$.  For our fermion problem, we have $\H = \H_F$ and 
\begin{align}
    & \betaA = u \c{0} - ve^{i \phi} \c{1}^\dagger,
    \label{eqn:jump}
\end{align}
where $u = \sqrt{1-v^2}$, with the pairing parameter $v$ real and satisfying $0\leq v\leq 1$, and $\Gamma$ parametrizes the strength of the dissipation, and corresponds to the cooling rate of the localized Bogoliubov mode $ \betaA$. The dissipation in \cref{eqn:master} induces an effective non-Hermitian Hamiltonian which includes pairing terms of the form $(iuve^{i \phi} c_{\1}^\dagger c_{\0}^\dagger - \hc)$.  Our system thus has the form of an unusual dissipative impurity problem, where the ``impurity" corresponds to the local dissipative pairing terms.  At a heuristic level, the dissipation injects Cooper pairs on these sites, which can then propagate outward in the lattice. Generically, \cref{eqn:master} will lead to an impure steady state, with fluxes of Cooper pairs both into and out of the lattice. We note that quadratic fermionic models with dissipative pairing have been studied previously in the context of cold atoms \cite{Diehl2010,Diehl2011,Iemini2016,Yi2012}, but unlike our work, these assumed pairing on every lattice site.

We next show that if the lattice Hamiltonian obeys a ubiquitous kind of generalized chiral symmetry, then we can ensure the existence of a unique, pure, entangled steady state \cite{Supplement}. This includes, but is certainly not limited to, lattices that can be divided into two equal sized sublattices, denoted $A$ and $B$, such that $\H_F$ only permits hopping from $A\leftrightarrow B$. This structure is found in many lattice systems, including all nearest neighbor hopping models on square lattices. Observe that this implies that $\H_F$ has a chiral symmetry, since the symmetry that sends $\hat c_i \to -\hat c_i$ if $ i \in A$ and $\hat c_i \to \hat c_i$ if $ i \in B$ sends $\H_F \to -\H_F$. This guarantees we can diagonalize the Hamiltonian such that $\H_F = \sum_{\alpha > 0} \epsilon_\alpha \left( \hat d_\alpha^\dagger \hat d_\alpha - \hat d_{-\alpha}^\dagger \hat d_{-\alpha} \right)$ where $\hat d_{\pm \alpha}^\dagger$ creates an eigenmode with energy $\pm \epsilon_\alpha$.

We can use this to rewrite the jump operator as a sum over non-local Bogoliubov modes, which pair positive and negative energy eigenmodes whenever the sites $\0,\1$ live on different sublattices
\cite{Supplement}:
\begin{align}
    \betaA &= \sum_\alpha N_\alpha (\hat \beta_\alpha + \hat \beta_{-\alpha}), \label{eqn:jump2} 
\end{align}
where 
\begin{align}
    \hat \beta_\alpha &= u_\alpha \hat d_\alpha - v_\alpha \hat d_{-\alpha}^\dagger \ \ \ \ \
    \hat \beta_{-\alpha} = u_\alpha \hat d_{-\alpha} + v_\alpha \hat d_{\alpha}^\dagger.
\label{eq:bogmodes}
\end{align}
These are an independent set of fermionic annihilation operators obeying canonical anticommutation relations. The constants $u_\alpha$ and $v_\alpha$ encode information about the overlap of the eigenmodes with the dissipation sites $\0,\1$ \cite{Supplement}:
\begin{align}
    u_\alpha &= \frac{u \psi_\alpha[\0]}{N_\alpha}, \ \ \ \ \ \
    v_\alpha = \frac{v e^{i \phi} \psi_{-\alpha}^*[\1]}{N_\alpha}, \\
    N_\alpha &= \sqrt{u^2 |\psi_\alpha[\0]|^2 + v^2 |\psi_{-\alpha}^*[\1]|^2},
\end{align}
where $N_\alpha$ fixes normalization.
For certain finely-tuned  parameters, it is possible that some $N_\alpha = 0$, in which case those eigenmodes have no overlap with the dissipation sites, and thus are not cooled. However, for a generic Hamiltonian, $N_\alpha \neq 0$, and so the jump operator is a linear combination of all $2N$ of the Bogliubov modes; the steady state is \textit{uniquely} their joint vacuum. This state is pure, and has entanglement that grows linearly with system size. We can express the steady state in terms of the eigenmodes as $|\psi\rangle = \prod_{\alpha>0} \left(  u_\alpha - v_\alpha \hat d_{-\alpha}^\dagger \hat d_\alpha^\dagger \right)|0\rangle$. The correlators in the steady state are:
\begin{align}
    \langle \hat d_\alpha^\dagger \hat d_\beta \rangle &= |v_\alpha|^2 \delta_{\alpha \beta}, \ \ \ \ \langle \hat d_\alpha \hat d_\beta \rangle = -u_\alpha v_\beta \sgn(\alpha) \delta_{\alpha, -\beta}. \label{eq:fermiCor}
\end{align}

In real space, the entanglement structure of our dissipatively stabilized steady state is only between the two sublattices $A$ and $B$, and the amount of entanglement between the two sublattices grows linearly with $N$ \cite{Supplement}. However, despite only being between the sublattices, the spatial pattern can be quite complicated. Given a generic Hamiltonian, any given lattice site will be correlated with the entire sublattice it does not reside on.
There are many systems that possess the chiral symmetry required for our scheme.  A particularly simple example (that, surprisingly, will generalize to the case of spins) is a 1D lattice with $2N$ sites described by a Hamiltonian with nearest neighbour hopping that possesses an inversion symmetry about its midpoint. Since $\H_F$ is symmetric, if the sites $\0,\1$ are mapped to each other by the mirror symmetry, then $u_\alpha = u, v_\alpha = v \ \forall \alpha$. The correlators defined in \cref{eq:fermiCor} are then also particularly simple in real space, giving the unique steady state $|\psi\rangle = \prod_{i=1}^N \left(  u - v(-1)^i \hat c_{-i}^\dagger \hat c_i^\dagger \right)|0\rangle$. This state exhibits volume-law entanglement (see \cref{fig:rainbow}), and is known as a rainbow state \cite{Vitagliano2010,Zhang2017}.  As discussed in \cite{Supplement}, many more examples are possible, including the 1D Su-Schrieffer-Heeger model \cite{Su1979,Su1980} and the 2D Hofstadter model \cite{Hofstader1976}.  


{\textit{Qubits-- }}
Using the above fermionic setup as inspiration, we now ask whether a similar dissipative preparation scheme is possible for an array of coupled spins or qubits. While any quadratic $A \leftrightarrow B$ hopping Hamiltonian worked for the fermions, it turns out the qubits require the slightly stricter condition of a 1D nearest-neighbor hopping chain. In this case, the sublattices are comprised of every-other lattice sites. Further, while the fermions would take any placement choice of $\0,\1$ so long as they were on different sublattices, the qubits require they be neighboring \cite{Supplement}. 

This still leaves all 1D nearest neighbor hopping models, including disordered systems. For simplicity, we will focus below on the case of a lattice with mirror symmetry. It was in this case that the fermions had simplified spatial correlations, giving rise to the rainbow structure. The master equation is then given by \cref{eqn:master} with $\H  = \H_S$ and   
\begin{align}
    \H_S  &= -\left[ \ \ \smashoperator{\sum_{i=-N}^{-2}} J_i \hat \sigma^+_i \hat \sigma^-_{i + 1}  + \smashoperator{\sum_{i=1}^{N-1}} J_i \hat \sigma^+_i \hat \sigma^-_{i + 1} +  J_{-1} \hat \sigma^+_{-1} \hat \sigma^-_{1} \right] \nonumber \\ & \ \ \ \ \ \ \ \ \ \ \ + \hc,
    \label{eq:Hspin} \\ 
    \betaA &= u\hat \sigma^-_{\0} + v\hat \sigma^+_{\1}. \label{Eq:SpinME}
\end{align}
Here $\hat{\sigma}^+_i$ ($\hat{\sigma}^-_i$) is the Pauli raising (lowering) operator on site $i$. Our lattice has $2N$ sites labeled $(-N,\dots,-1,1,\dots N)$, i.e. there is no $0^{\text{th}}$ lattice site. We will constrain the hoppings to obey $J_i= J_{-i-1}$, so $\H$ has mirror symmetry. The dissipation-coupled sites $\0$ and $\1$ will then be the middle two sites of the lattice, i.e.~$\0 = -1, \1 = 1$.  While the dissipator here may seem exotic, we show below how they can be realized in a number of platforms using existing experimental tools.    

The above spin model can be readily mapped to fermions using the Jordan-Wigner (JW) transformation \cite{Jordan1928}.  However, it necessarily maps to an {\it interacting} fermionic model (in contrast to the quadratic system considered in \cref{eqn:jump}).  The most convenient mapping to JW fermions $\hat{c}_j$ is given by the transformation    
\begin{align}
\hat{c}_i = 
\begin{cases}
\left(\prod_{j = 1}^i \hat \sigma_j^z \right) \hat  \sigma_i^-
& 1 \leq i \leq N  
, \\
\left(\prod_{j = 1}^N \hat \sigma_j^z \right)\left(\prod_{j = -N}^i \hat \sigma_j^z \right) \hat \sigma_i^-
& -N \leq i \leq -1
\label{eq:JW.center}
.
\end{cases} 
\end{align}
This corresponds to using site $1$ as the reference for the string operators.
Letting $\hat{N}_{\rm tot}$ be the total fermion number operator, our model can be expressed in terms of these fermionic degrees of freedom as: 
\begin{align}
   & \H_S = 
   \displaystyle\smashoperator{\sum_{i \neq N,-1}} J_i \hat{c}_i^\dagger \hat{c}_{i + 1} + J_{-1}(-1)^{\hat{N}_{\rm tot}} \hat{c}^\dagger_{1} \hat{c}_{-1}
   + \hc 
            \label{eq:newGauge}
     ,   \\
    & \betaA = u \hat{c}_{-1} (-1)^{\hat{N}_{\rm tot}} - v \hat{c}^\dagger_1. 
\end{align}


\begin{figure*}[t!]
\centering
\begin{subfigure}
    \centering
    \includegraphics[width = 3.25in]{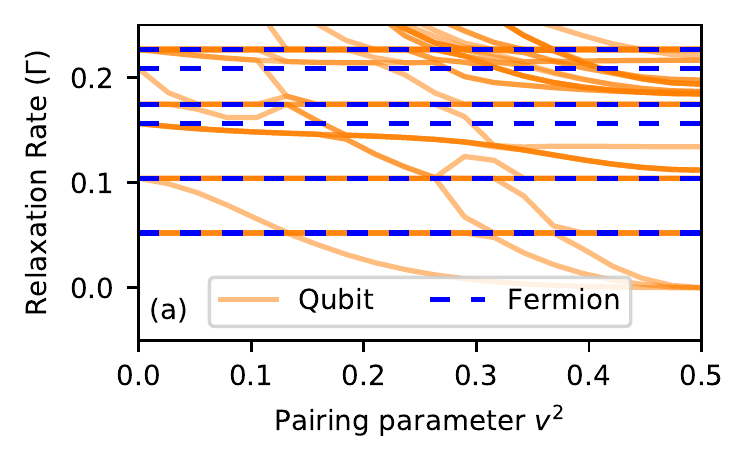}
\end{subfigure}
\begin{subfigure}
    \centering
    \includegraphics[width = 3.4in]{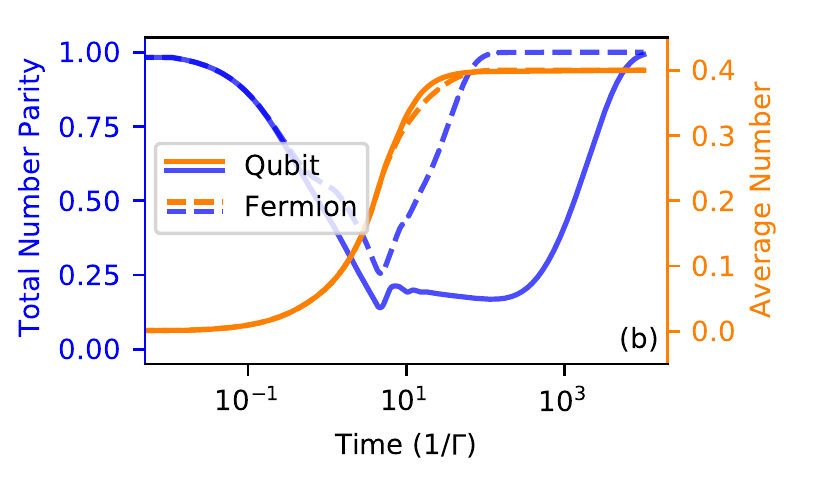}
\end{subfigure}
\caption{
(a)  Low-lying dissipative spectrum of our master equation in \cref{eqn:master} (i.e.~real part of Liouvillian eigenvalues) for a $6$ site 1D lattice with $\Gamma = J_j$, plotted as a function of $v^2$. We take the simple case where the Hamiltonians $\H_F, \H_S$ have a mirror symmetry. Both the qubit model (orange) and the free-fermion model (blue) are shown.  While the spectra coincide for the trivial $v=0$ case, the fermionic spectrum is independent of $v$, whereas for qubits, slow modes and complex level structure arise as $v$ is increased.  (b)  Time evolution of both the average excitation number (orange) and total number parity $(-1)^{\hat{N}}$ (blue) for an $8$-site lattice, $v^2=0.4$, starting from the vacuum 
state; other parameters same as (a).  The qubit model exhibits an extremely slow relaxation of total number parity, a direct reflection of the pairing phase fluctuations that emerge in its fermionic representation.}     
\label{fig:dynamics}
\end{figure*}


We see that the presence of the phase operator $(-1)^{\hat{N}_{\rm tot}}$ in both $\H_S$ and the dissipative terms ruins a mapping to free fermions.  On a heuristic level, we can interpret this as a modification of \cref{eqn:jump} that now describes fluctuations in the phases of the Cooper pairs injected into the system by the reservoir.  
For the simple fermionic system described by \cref{eqn:jump}, pairs are always injected with a fixed phase $\phi$; in contrast, in \cref{eq:newGauge}, they are injected with a phase $\pm 1$ that depends on the system's parity.  We stress that even with other gauge conventions for the JW transformation, it is not possible to eliminate these phase fluctuations (i.e.~string operators) from the dissipator.  

To better understand our system, we can re-write the nonlinear dissipation operators in terms of a fixed basis of local Bogoliubov operators $\hat{\bar{\beta}}_A = u \hat c_{-1} - v \hat c_1^\dagger, \hat{\bar{\beta}}_B = u \hat c_1 + v \hat c_{-1}^\dagger$. As was noted in the preceding section, we can write $\hat{\bar{\beta}}_A$ as a sum over the modes defined in \cref{eq:bogmodes}. Further, in the special case of a mirror symmetric Hamiltonian $\hat{\bar{\beta}}_B$ will also be a sum over these modes, \cite{Supplement}. Defining $\hat{P}_{\rm ev} = (1 + (-1)^{\hat{N}_{\rm tot}})/2$ as the projection operator onto even number-parity states, we have:  
\begin{align}
    \betaA & = \hat{\bar{\beta}}_A \hat{P}_{\rm ev} - 
        \left[ (u^2 - v^2)\hat{\bar{\beta}}_A + 2uv\hat{\bar{\beta}}_B^\dagger \right] (1 - \hat{P}_{\rm ev}) .
        \label{eq:BetasParity}
\end{align}
This provides a simple way to understand the phase fluctuation physics: it is as though the dissipation has a parity-dependent temperature.  For even-parity states, the dissipation can only remove Bogoliubov exctiations, i.e.~it acts like an effective zero temperature bath.  In contrast, for odd-parity states and $v\neq0$, we see that there are amplitudes for the dissipation to either create or destroy excitations (like an effective bath at a non-zero temperature).   

\cref{eq:BetasParity} also leads to an important conclusion:  despite the additional nonlinearity and phase fluctuations in our spin model, the steady state of our simple free fermion model in \cref{eqn:jump} is also a steady state of the spin model.  This steady state (which here is a rainbow state, given the mirror symmetry of $\H$) has a definite even number parity, and hence the Liouvillian acting on this state is identical to the free fermion Liouvillian.  At a heuristic level, this state has a definite number parity, and hence phase-fluctuations are irrelevant.  Returning to our original qubit degrees of freedom, the pure steady state takes the form:
\begin{align}
    |\psi_{\rm ss} \rangle = \prod_{i = 1}^N (u + (-1)^i v\hat \sigma_i^+ \hat \sigma^+_{-i})|0\rangle.  \label{eq:SpinSS}
\end{align}
\cref{eq:BetasParity} also lets us show that as long as $v^2 \neq 1/2$, this steady state is unique \cite{Supplement}. Since the Hamiltonian conserves Bogoliubov excitations, moving between manifolds with different Bogoliubov number can only happen dissipatively. The state with no Bogoliubov excitations has even parity, so it can only be cooled by \cref{eq:BetasParity}; however, there is no lower number state, and so it is dark to the dissipation. On the other hand, every higher excitation state can be cooled, and so eventually all of the population flows to the Bogoliubov vacuum. Note, this argument also holds for non-rainbow symmetric systems \cite{Supplement}. Thus, we have a central result of this Letter:  any initial state of our qubit array (irrespective of its purity or entanglement) will relax into this volume-law entangled pure state. 
Further, this result holds independently of the magnitude of the hopping parameters $J_j / \Gamma$, and even in the presence of additional Hamiltonian terms that preserve the mirror symmetry of $\H_S$ \cite{Supplement}.

We can now also understand the additional constraints put on the qubits to maintain the purity of the steady state. The fact that the chain must be 1D nearest neighbor hopping is a result of the non-local nature of the Jordan-Wigner Transformation. In the same vein, requiring the dissipation coupled sites to be neighboring tells us the only string operator that appears in the master equation is $(-1)^{\hat{N}_\text{tot}}$. Despite these constraints, we can see that our model is still incredibly robust to disorder. The steady state will still be pure, unique and entangled regardless of the different hopping amplitudes between lattice sites. See \cite{Supplement} for more details.

{\it Dynamics and multi-stability-- } While the qubit and free-fermion dissipative arrays share the same pure steady state, the models have strikingly different dynamics.  This is a direct consequence of the form of the dissipator given in \cref{eq:BetasParity}.  The phase fluctuations in the fermionic representation of the qubit model lead to slower overall relaxation,  due to the effective non-zero temperature and excitation-creation associated with odd-parity states.  As can be seen from \cref{eq:BetasParity}, this odd parity heating increases as $v$ is increased from 0, with an amplitude $\propto v \sqrt{1-v^2}$.  For free fermions, there is no heating:   the dissipative dynamics always corresponds to removing excitations, irrespective of the system state.  

Shown in \cref{fig:dynamics}a is the numerically-calculated dissipative spectrum of the Liouvillians for the free fermion and qubit versions of our model as a function of the pairing parameter $v^2$.  For free fermions, the relaxation rates are independent of $v^2$;
one can show that for large $N$, the slowest relaxation rate (dissipative gap) scales as $1/N^3$ \cite{Supplement}.  In stark contrast, the relaxation rates in the qubit model depend on $v^2$, with the emergence of an extremely small dissipative gap as $v^2 \rightarrow 1/2$.  \cref{fig:dynamics}b demonstrates that this emergent slow timescale manifests itself directly in observable quantities.  We see that for both the qubit and free-fermion models, the average particle numbers relax on a similar timescale $\sim 1/\Gamma$.  The average parity relaxes on the same timescale for fermions, but for the qubit model, exhibits exponentially slower relaxation.  
This is a direct manifestation of the effective phase fluctuations encoded in \cref{eq:newGauge}.     

The case $u^2 = v^2 = 1/2$ is also of special interest.  \cref{eq:BetasParity} indicates that in this case, the dissipation can only remove excitations from even parity states, and can only add excitations to odd parity states.  This immediately leads to multi-stability, as if the system starts in a state with $2m$ Bogoliubov excitations, it will forever be stuck in a manifold of states having either $2m$ or $2m-1$ excitations.  This immediately leads to at least $N+1$ steady states (see \cite{Supplement} for more details).  We stress that there is no multi-stability in the free-fermion model.  


{\textit{Experimental Implementation --}}  The basic qubit master equation in \cref{eq:Hspin,Eq:SpinME} could be realized in a variety of platforms.  Linear arrays of tunnel-coupled qubits have been realized in many systems, including trapped ions 
\cite{MonroeNature2014,BrydgesScience2019,KokailZollerNature2019} and superconducting qubits \cite{Roushan2017,HouckPRX2017,Ma2019}.  The required dissipation on sites $j=-1,1$ could be achieved by driving these qubits with two-mode squeezed light via transmission lines or waveguides \cite{Kraus2004}.  In this case, $\Gamma$ would represent the waveguide coupling rates, and $v/u = \tanh r$, with $r$ the squeezing parameter.  While such a scheme could be realized by driving superconducting qubits with two-mode squeezed microwave radiation generated by a Josephson parametric amplifier \cite{Eichler2011,Huard2012}, implementation routes that do not require non-classical microwaves or light are also possible.  The required dissipator can be realized by coupling qubits $j=-1,1$ to a common dissipative bosonic mode (e.g.~a lossy microwave cavity), and then either modulating the qubit frequencies, or modulating the qubit-resonator couplings (as was recently achieved \cite{SchusterPRL2017}).  By interfering e.g.~a red sideband process on one qubit with a blue sideband process on the second qubit, the required dissipator can be achieved (with $u,v$ being determined by the modulation amplitudes). Interfering red and blue sideband processes has been used previously in both trapped ion \cite{Kienzler2015} and superconducting qubit \cite{HuardPRX2021} experiments for reservoir engineering bosonic modes, but not to control qubit dissipation in the way we suggest.  More details on this modulation approach, and on the resilience of our scheme to disorder and unwanted dissipation (i.e.~qubit dephasing and relaxation) are presented in \cite{Supplement}.


{\textit{Conclusions --}}  We have demonstrated that the combination of spatially-localized pairing dissipation with symmetry contrained Hamiltonian dynamics can be used generically to stabilize entangled states in systems with locally-constrained Hilbert spaces.  These states can exhibit long range, volume-law entanglement.  In the case of a qubit array, our setup corresponds to a dissipative spin chain that is equivalent to an interacting fermionic model, which can be interpreted in terms of dissipative Cooper pairing with phase fluctuations.  Our ideas are compatible with a number of different experimental platforms, and could provide an important resource for a variety of quantum information processing protocols.   


This work is supported by the Air Force Office of Scientific Research MURI program under Grant No. FA9550- 19-1-0399.

\nocite{Ma2017,Zippilli2021,Yanay2018,Buca2012,Jaynes1963,Shore1993,Rempe1987,Gardiner2004,Gardiner1985,SchusterPRL2017,Ma2019}

\bibliography{ref}
\end{document}


\title{Supplemental Material: Stabilizing volume-law entangled states of fermions and qubits using local dissipation}

\author{Andrew Pocklington}
\affiliation{Pritzker School of Molecular Engineering,  University  of  Chicago, 5640  South  Ellis  Avenue,  Chicago,  Illinois  60637,  U.S.A.}

\author{Yu-Xin Wang}
\affiliation{Pritzker School of Molecular Engineering,  University  of  Chicago, 5640  South  Ellis  Avenue,  Chicago,  Illinois  60637,  U.S.A.}

\author{Yariv Yanay}
\affiliation{Laboratory for Physical Sciences, 8050 Greenmead Dr., College Park, MD 20740}

\author{A. A. Clerk}
\affiliation{Pritzker School of Molecular Engineering,  University  of  Chicago, 5640  South  Ellis  Avenue,  Chicago,  Illinois  60637,  U.S.A.}

\date{\today}

\maketitle

\renewcommand{\theequation}{S\arabic{equation}}
\renewcommand{\thesection}{\Roman{section}}
\renewcommand{\thefigure}{S\arabic{figure}}
\renewcommand{\thetable}{S\arabic{table}}
\renewcommand{\bibnumfmt}[1]{[S#1]}
\renewcommand{\citenumfont}[1]{S#1}


\bigskip

\section{Fermions}
\label{SIsec:1}

In the main text, we show that when equipped with the appropriate chiral symmetry, the fermionic master equation has a unique, pure steady state. In this section, we provide a detailed proof for this result. Recall the Hamiltonian and jump operators:
\begin{align}
&\H_F = \sum_{i,j = 1}^{2 N} H_{ij} \hat{c}_i^\dagger \hat{c}_j  
    = \smashoperator{\sum_{\alpha = 1 }^{  N}}  \epsilon_\alpha (\hat{d}_\alpha^\dagger \hat{d}_\alpha - \hat{d}_{-\alpha}^\dagger \hat{d}_{-\alpha}) 
    , \\
   &  \betaA = u \c{0} - ve^{i \phi} \c{1}^\dagger.
\end{align}
Our strategy is to find a new set of Bogoliubov energy eigenmodes so that the dissipator $\hat \beta_L$ cools these modes into their joint vacuum in the steady state limit. In the following, we first show that the Hamiltonian is invariant under a set of Bogoliubov transformations if it has a built-in chiral symmetry, and then discuss conditions for the system to have the desired pure steady state.


\subsection{General Model}
\label{SIsec:1A}

We assume that our system has $2N$ sites, and a chiral symmetry, $\C$, such that $\C \H_F \C^\dagger = -\H_F$. This means there are $2N$ eigenmodes, and they come in pairs of energy $\pm \epsilon_\alpha$, where $\C \hat d_\alpha \C^{-1} = \hat d_{-\alpha}$. Hence, we can label our eigenmodes $\hat d_{\pm \alpha}^\dagger$, where $\alpha = 1, \dots, N$, and $d_{\pm \alpha}^\dagger$ creates a mode with energy $\pm \epsilon_\alpha$. We will define the unitary transformation $\hat c_i = \sum_\alpha \psi_\alpha[i] \hat d_\alpha$ to go between position space and energy eigenmode space. We will posit that our system has the steady state solution $|\psi\rangle_{ss}$ of the form
\begin{align}
    |\psi\rangle_{ss} &= \prod_{\alpha > 0} (u_\alpha + v_\alpha d_\alpha^\dagger d_{-\alpha}^\dagger)|0\rangle, \label{SIeqn:steadystate}
\end{align}
where $|0\rangle$ is the joint vacuum of all of the energy eigenmodes, and $u_\alpha, v_\alpha \in \CC$ are yet to be determined constants such that $|u_\alpha|^2 + |v_\alpha|^2 = 1$, to maintain normalization. It is evident by construction that this is a zero energy eigenstate of the Hamiltonian. We can now observe that
\begin{align}
    \betaA  |\psi\rangle_{ss} &= (u \c{0} - ve^{i \phi} \c{1}^\dagger) \prod_{\alpha > 0} (u_\alpha + v_\alpha d_\alpha^\dagger d_{-\alpha}^\dagger)|0\rangle \\
    &= \sum_{\gamma} (u \psi_\gamma[\0] \hat d_\gamma - ve^{i \phi} \psi^*_\gamma[\1] \hat d_\gamma^\dagger) \prod_{\alpha > 0}  (u_\alpha + v_\alpha d_\alpha^\dagger d_{-\alpha}^\dagger)|0\rangle \\
    &= \sum_{\gamma>0} \prod_{\alpha \neq \gamma} (u_\alpha + v_\alpha d_\alpha^\dagger d_{-\alpha}^\dagger)  (u \psi_\gamma[\0] \hat d_\gamma + u \psi_{-\gamma}[\0] \hat d_{-\gamma} - ve^{i \phi} \psi^*_\gamma[\1] \hat d_\gamma^\dagger - ve^{i \phi} \psi^*_{-\gamma}[\1] \hat d_{-\gamma}^\dagger) (u_\gamma + v_\gamma d_\gamma^\dagger d_{-\gamma}^\dagger)|0\rangle \\
    &= \sum_{\gamma>0} \prod_{\alpha \neq \gamma} (u_\alpha + v_\alpha d_\alpha^\dagger d_{-\alpha}^\dagger)  (u v_\gamma \psi_\gamma[\0] \hat d_{-\gamma}^\dagger - u v_\gamma \psi_{-\gamma}[\0] \hat d_{\gamma}^\dagger - u_\gamma ve^{i \phi} \psi^*_\gamma[\1] \hat d_\gamma^\dagger - u_\gamma ve^{i \phi} \psi^*_{-\gamma}[\1] \hat d_{-\gamma}^\dagger)|0\rangle \\
    &= \sum_{\gamma>0} \prod_{\alpha \neq \gamma} (u_\alpha + v_\alpha d_\alpha^\dagger d_{-\alpha}^\dagger)  \left[ (u v_\gamma \psi_\gamma[\0] - u_\gamma ve^{i \phi} \psi^*_{-\gamma}[\1] ) \hat d_{-\gamma}^\dagger - (u v_\gamma \psi_{-\gamma}[\0]  + u_\gamma ve^{i \phi} \psi^*_\gamma[\1]) \hat d_\gamma^\dagger \right] |0\rangle.
\end{align}
If this is a steady state, then it must be zero. Hence, we will require that 
\begin{align}
    u v_\gamma \psi_\gamma[\0] - u_\gamma ve^{i \phi} \psi^*_{-\gamma}[\1] &= u v_\gamma \psi_{-\gamma}[\0]  + u_\gamma ve^{i \phi} \psi^*_\gamma[\1] = 0 \label{Seq:constraint1}.
\end{align}
This equation will have solutions if, and only if
\begin{align}
    \frac{\psi_\gamma[\0]}{\psi_{-\gamma}[\0]} &= -\frac{\psi^*_{-\gamma}[\1]}{\psi^*_\gamma[\1]} \label{Seq:constraint2}.
\end{align}
This sets a constraint on both the placement of $\0,\1$, as well as on the overall chiral symmetry. Assuming \cref{Seq:constraint2} holds, we can define 
\begin{align}
    u_\gamma &= \frac{u \psi_\gamma[\0]}{N_\gamma} \\
    v_\gamma &= \frac{v e^{i \phi} \psi_{-\gamma}^*[\1]}{N_\gamma}, \\
    N_\gamma &= \sqrt{u^2 |\psi_\gamma[\0]|^2 + v^2 |\psi_{-\gamma}^*[\1]|^2},
\end{align}
where $N_\gamma$ enforces normalization. These allow us to define the Bogoliubov modes
\begin{align}
    \hat \beta_\gamma &= u_\gamma \hat d_\gamma - v_\gamma \hat d_{-\gamma}^\dagger, \\
    \hat \beta_{-\gamma} &= u_\gamma \hat d_{-\gamma} + v_\gamma \hat d_{\gamma}^\dagger.
\end{align}
We can then rewrite the Hamiltonian nicely as
\begin{align}
    \H_F &= \sum_{\gamma > 0}\epsilon_\gamma (\hat \beta_\gamma^\dagger \hat \beta_\gamma - \hat \beta_{-\gamma}^\dagger \hat \beta_{-\gamma}). \label{SIeqn:BogHam}
\end{align}
Note that $\hat \beta_{\pm \gamma} |\psi \rangle_{ss} = 0$, meaning the steady state is the joint vacuum of all of the Bogoliubov modes, and also a 0 energy eigenstate. We can now observe that
\begin{align}
    \betaA &= u\c{0} - v e^{i\phi}\c{1}^\dagger \\
    &= \sum_{\gamma} u \psi_\gamma[\0] \hat d_{\gamma} - v e^{i\phi} \psi_\gamma^*[\1] d_\gamma^\dagger \\
    &= \sum_{\gamma > 0} u \psi_\gamma[\0] \hat d_{\gamma} + u \psi_{-\gamma}[\0] \hat d_{-\gamma} - v e^{i\phi} \psi_\gamma^*[\1] d_\gamma^\dagger - v e^{i\phi} \psi_{-\gamma}^*[\1] d_{-\gamma}^\dagger \\
    &=  \sum_{\gamma > 0} N_\gamma \left[  u_\gamma \hat d_{\gamma} + u_\gamma x_\gamma^{-1} \hat d_{-\gamma} + x_\gamma^{-1} v_\gamma d_\gamma^\dagger - v_\gamma d_{-\gamma}^\dagger \right] \\
    &=  \sum_{\gamma > 0} N_\gamma \left[  \hat \beta_\gamma + x_\gamma^{-1} \hat \beta_{-\gamma} \right],
\end{align}
where we have defined $x_\gamma = \psi_\gamma[\0]/\psi_{-\gamma}[\0]$, the ratio in \cref{Seq:constraint2}. For an $A \leftrightarrow B$ hopping model, we can work in a gauge where $x_\gamma = 1$, reproducing the result in the main text. Thus, we can see that the jump operator $\betaA$ cools all of the Bogoliubov modes with a rate fixed by the overlap of the eigenmodes with the dissipation sites, encoded in $N_\gamma$. Thus, the steady state will always be uniquely the joint vacuum of the modes $\hat \beta_\gamma$, whenever there are no eigenmodes dark to the dissipation sites. The original Hamiltonian had a $U(1)^{2N}$ gauge symmetry, which came from rotating the phases individually of any of the eigenmodes. In a very specific way, the jump operator has broken that gauge freedom to be just $U(1)^N$, as now the relative phase between the positive and negative eigenmodes is fixed in the definition of the Bogoliubov modes. This dissipative symmetry breaking is also apparent in \cref{Seq:constraint1}, which can only be satisfied for a fixed phase reference between positive and negative eigenmodes. The steady state is Gaussian, and can be characterized by its correlation matrices. In energy eigenmode space, it is simply
\begin{align}
\langle \hat d_\alpha^\dagger \hat d_\gamma \rangle &= |v_\alpha|^2 \delta_{\alpha \gamma}, \ \ \ \ \ \  \langle \hat d_\alpha \hat d_\gamma \rangle = u_\alpha v_\gamma \sgn(\alpha) \delta_{\alpha, -\gamma}, \label{Seq:correlator}
\end{align}
where $\sgn$ is the sign function.

Finally, we can immediately observe that the decay rates are set by $N_\gamma$, which depends on the overlap of the eigenmodes with the dissipation sites. The dissipative gap, or the slowest decay rate, is therefore determined by $\min_\gamma |N_\gamma|^2$. Normalization tells us this scales like at best $1/N^2$ when the modes are approximately evenly distributed across the lattice, and possibly much worse if the modes are highly localized.

\subsection{$A \leftrightarrow B$ Hopping Models}
\label{SIsec:1B}

A very simple, and general model that satisfies the condition in \cref{Seq:constraint2} is a lattice that can be divided into two, equal size sublattices, denoted $A,B$, such that the Hamiltonian $\H_F$ only allows hopping between the sublattices. For example, any nearest neighbor hopping model on a square lattice can be described this way, where the sublattices are comprised of every other lattice site (or a checkerboard-like pattern in higher dimensions). Now, in this case, we have the chiral symmetry, as mentioned in the main text:
\begin{align}
    \C: \hat c_i \to \left\{ \begin{array}{cc}
     -\hat c_i   &  i \in A\\
    \hat c_i     & i \in B
    \end{array}  \right. ,
\end{align}
which sends the Hamiltonian $\H_F \to -\H_F$.
Now, if we let $\0 \in A, \ \ \ \1 \in B$, then we can observe that $\psi_\gamma[\0] = -\psi_{-\gamma}[\0]$ and $\psi_\gamma[\1] = \psi_{-\gamma}[\1]$, immediately satisfying the constraint \cref{Seq:constraint2}. We can rewrite the Hamiltonian, as a matrix equation, by defining $\vec{c} = (\hat c_1, \dots \hat c_{2N})^T$ where sites $1,\dots,N\in A$ and $N +1, \dots 2N \in B$, to get
\begin{align}
    \H_F &= (\vec{c} \ )^\dagger \left( \begin{array}{cc}
        0 & V \\
        V^\dagger & 0
    \end{array}\right) \vec{c},
\end{align}
with $V$ some generic matrix encoding the hopping. One chiral symmetry, $\C$, is written above, which simply sends $A \to -A$ and $B \to B$. However, any Hamiltonian with 1 chiral symmetry in fact has an infinite $U(1)^{2N}$ family of chiral symmetries, since we can send $\hat d_\alpha \mapsto e^{i \phi_\alpha} \hat  d_{-\alpha}$ where the phase is a free, $U(1)$ gauge symmetry for each eigenmode. Now, the correlators in \cref{Seq:correlator} tell us the preferred chiral symmetry should be antisymmetric, as is required by fermionic statistics. Hence, we can define a new chiral symmetry $\hat{S}:\hat d_\alpha \mapsto \sgn(\alpha) \hat  d_{-\alpha}$. If we now express this in real space, just like the Hamiltonian, we get that
\begin{align}
    S &= \left( \begin{array}{cc}
        0 & U \\
        -U^\dagger & 0
    \end{array}\right) ,
\end{align}
where $U$ is a unitary $N \times N$ matrix. Now, since this a chiral symmetry, we can conclude that $U^\dagger V U^\dagger = V^\dagger$. This is equivalent to stating that $U$ is the unique, unitary matrix defined by the right polar decomposition of $V$. I.e., we can rewrite $V$ uniquely as $V = WU$ where $W$ is Hermitian. We should think of $U$ as being a map of the eigenmodes of $V$, which live on the $A$ lattice, to the eigenmodes of $V^\dagger$, which live on the $B$ lattice, and $U^\dagger$ begin the inverse mapping. Because $V,V^\dagger$ have the same spectrum, it makes sense to pair up equal energy eigenmodes. The steady state entanglement structure will then dissipatively pair the two sets of eigenmodes completely through the matrix $U$. The Hermitian part, $W$, sets the spectrum, and also determines the constants $u_\alpha, v_\alpha$. 

Now, in the case that $V = V^\dagger$, this reduces to the case in the text where we have a mirror symmetry of the lattice. Hence, we know $U = 1$ and so we can immediately read off that the entanglement structure will be the identity, pairing lattice sites reflected to each other by the mirror symmetry. We were able to read off this steady state without solving any of the dynamics, completely through symmetry arguments.

We can now calculate the log negativity, an entanglement monotone, to see that this does indeed obey a volume law. Here, we will define the new modes $\hat \gamma_\alpha^\pm = \frac{1}{\sqrt{2}}(\hat d_\alpha \pm \hat d_{-\alpha})$. These have the nice property of overlapping with only a single sublattice (and are in fact the eigenmodes of $VV^\dagger$ and $V^\dagger V$, respectively). We can now rewrite the steady state nicely as
\begin{align}
    |\psi\rangle_{ss} &= \prod_{\alpha > 0}(u_\alpha - v_\alpha (\hat \gamma_\alpha^+)^\dagger (\hat \gamma_\alpha^-)^\dagger)|0\rangle .
\end{align}
Now, it is very easy to see that all of the entanglement is between, and not within, the sublattices. The log negativity can now easily be calculated as
\begin{align}
    E_N &= \sum_{\alpha > 0} \log_2(1 + 2|u_\alpha v_\alpha|),
\end{align}
which clearly grows linearly with system size.


\subsection{Fermion Examples}
\label{SIsec:1C}

In the main text, we have given the fermionic example of the rainbow state, but more complicated, and higher dimensional lattice models can also create pure, entangled steady states. As mentioned in the main text, many exotic lattice models exhibit the necessary symmetry conditions, including the Hofstader lattice with a quarter flux and the Su-Schriefer-Heeger model, \cite{Yanay2018}.


\begin{figure}[H]
\centering
\begin{subfigure}
    \centering
    \includegraphics[width = 3in]{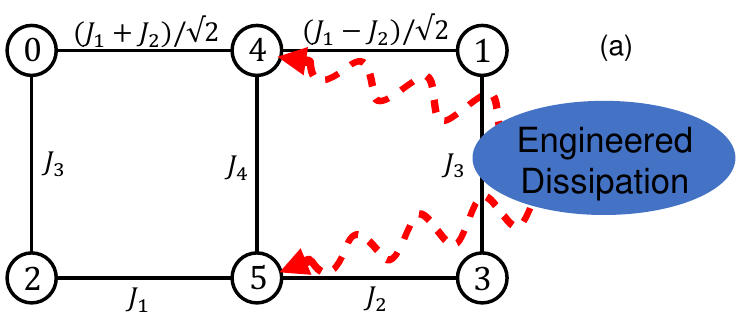}
\end{subfigure}
\begin{subfigure}
    \centering
    \includegraphics[width = 3in]{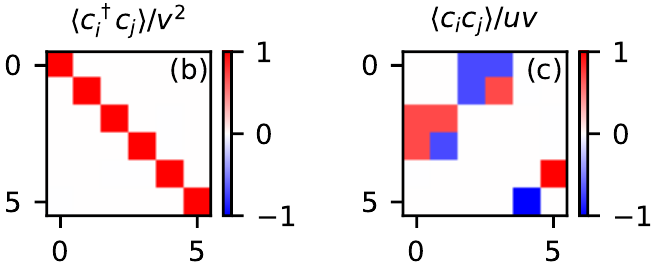}
\end{subfigure}
\caption{\footnotesize{(a) A 2D nearest neighbor hopping Hamiltonian with two sites coupled to an engineered reservoir. With certain constraints on the hopping amplitudes, the steady state is a pure entangled state with a non-rainbow form. Pictured are 7 possible bonds; $J_1,\dots,J_4$ are all free parameters, with the remaining 3 constrained by the symmetry conditions. (b) The correlators $\langle \hat{c}_i^\dagger c_j \rangle = v^2\delta_{ij}$ as expected. (c) The anomalous correlators show Bell state entanglement between the two sites coupled to the dissipation, but sites $0,1$, are entangled with superpositions of sites $2,3$.}}
\label{fig:nonRainbow}
\end{figure}


For one example of a more complex steady state that is not a rainbow, see \cref{fig:nonRainbow}. However, note that there is some constraint on the complexity of the correlations we can produce. Any steady state will always be a ``generalized'' rainbow state, in the sense that correlators always exactly swap two sublattice modes. In a true rainbow state, these modes are individual lattice sites, but they can be more complicated looking in real space \cite{Ma2017,Zippilli2021}.


\section{Qubit Steady State}
\label{SIsec:2}

In the main text, we have shown that a spin version of the general master equation also has a pure, unique steady state. There, we focused on the rainbow, but here we will derive the result in complete generality. Again using a master equation of the form of \cref{eqn:master}, the spin system has Hamiltonian and dissipators:
\begin{align}
&\H_S = - \sum_{i = 1}^{2N} J_i \hat \sigma^+_i \hat \sigma^-_{i + 1} + \hc \label{Eq:SpinHamSupp}
    , \\
&   \betaA =  u\hat \sigma^-_{k} - v \hat \sigma^+_{k + 1} \label{Eq:SpinJumpSupp}
    .
\end{align}
We take site $2N +1 \equiv 1$ and the spin counterpart of vacuum state $|0\rangle$ via the relation $ \hat \sigma^-_{k} |0\rangle \equiv 0 $. This gives open boundary conditions if $J_{2N} = 0$, and periodic boundary conditions otherwise. The site $k$ is arbitrary.

\subsection{Existence}
\label{SIsec:2A}

To calculate the steady state solution, we will transform to the fermion model via the Jordan-Wigner transformation (defining fermion number operators $  \hat{n}_{j} \equiv \hat{c}_{j} ^\dag \hat{c}_{j}$)
\begin{align}
\hat \sigma_i^- &= (-1)^{\sum_{j = k + 1}^i \hat n_j} \hat c_i,
\label{Seq:JW1}
\end{align}
The string operators are chosen such that the fermionic version of the Hamiltonian is quadratic everywhere but on the two sites connected to the dissipation:
\begin{align}
&\H_S =  \sum_{i \neq k} J_i \hat{c}_i^\dagger \hat{c}_{i + 1} - J_k \hat{c}_k^\dagger \hat{c}_{k + 1} (-1)^{\hat N} + \hc, \label{eq:fermionizedHam}
\end{align}
where $\hat N = \sum_i \hat n_i$ is the total number operator. The jump operator in \cref{Eq:SpinJumpSupp} then becomes
\begin{align}
    \betaA &= -(u \hat c_k (-1)^{\hat N} - v \hat c_{k + 1}^\dagger).
\end{align}
Since phases of Lindblad jump operators do not enter the dynamics, we will drop the overall minus sign henceforth.

Consider the state 
\begin{align}
    |\psi\rangle_{ss} &= \prod_{\alpha > 0} (u_\alpha + v_\alpha d_\alpha^\dagger d_{-\alpha}^\dagger)|0\rangle,
\end{align}
where $d_\alpha^\dagger$ are the energy eigenmodes of \cref{eq:fermionizedHam} in the even parity section, i.e. when we replace $(-1)^{\hat{N}}$ with $1$. Now, the first thing to observe is that $\hat N = \sum_i \hat c_i^\dagger \hat c_i = \sum_\alpha \hat d_\alpha^\dagger \hat d_\alpha$, since the energy eigenmodes $\hat d_\alpha$ are related to the real space operators by a unitary transformation. Further, since the fermionic steady state is completely paired, the parity operator $(-1)^{\hat N}$ will always have eigenvalue 1 acting on the steady state; i.e. $(-1)^{\hat N} |\psi\rangle_{ss} = |\psi\rangle_{ss}$. Thus, in the steady state, we can replace $(-1)^{\hat N}$ with 1, telling us the state identified above is trivially steady under the fermionic representation of the spin master equation in \cref{Eq:SpinHamSupp,Eq:SpinJumpSupp}.

\subsection{Uniqueness}
\label{SIsec:2B}

In the following, we will prove that the steady state identified above via the Jordan-Wigner transformation is also unique. To do this, we must first observe that the parity operator $(-1)^{\hat N}$ is a weak symmetry of the dynamics. Now, the Hamiltonian is number conserving, so it is clear that $[(-1)^{\hat N}, \H_S] = 0$. Parity anticommutes with $\hat c_i^{(\dagger)}$, so we can observe that
\begin{align}
    \D[\betaA]\left( (-1)^{\hat N} \hat \rho (-1)^{\hat N}\right)  &= \frac{1}{2}\left(2\betaA (-1)^{\hat N} \hat \rho (-1)^{\hat N} \betaA^\dagger - \betaA^\dagger \betaA (-1)^{\hat N} \hat \rho (-1)^{\hat N} - (-1)^{\hat N} \hat \rho (-1)^{\hat N} \betaA^\dagger \betaA \right) \\
    &= \frac{1}{2}\left(2(-1)^{\hat N}\betaA  \hat \rho\betaA^\dagger (-1)^{\hat N}  - (-1)^{\hat N}\betaA^\dagger \betaA  \hat \rho (-1)^{\hat N} - (-1)^{\hat N} \hat \rho  \betaA^\dagger \betaA(-1)^{\hat N} \right) \\
    &= (-1)^{\hat N} \left( \D[\betaA] \hat \rho \right) (-1)^{\hat N}
    . 
\end{align}
This, along with commuting with the Hamiltonian, gives us 
\begin{align}
    \L[(-1)^{\hat N} \hat \rho (-1)^{\hat N}] &= (-1)^{\hat N} \L[\hat \rho ] (-1)^{\hat N}
    , 
\end{align}
which is the definition of a weak symmetry. This means that in the steady state, there are no coherences between positive and negative parity states \cite{Buca2012}, and so it makes sense to decompose this into two different systems - one with $ (-1)^{\hat N} = 1$ (even parity) and one with $ (-1)^{\hat N} = -1$ (odd parity). 

If we define $\Pev$ as the projector into the even parity subspace, we know that by the chiral symmetry arguments in \cref{SIsec:1} (\cref{SIeqn:BogHam}), we can rewrite the even-parity Hamiltonian $\Pev \H_S \Pev $ as a sum over the Bogoliubov modes $ \sum_\alpha \epsilon_\alpha \hat{\beta}_\alpha^\dagger \hat{\beta}_\alpha$, meaning it conserves these modes. Next, define $\hat{\bar \beta}_A = u\hat{c}_{k} - v\hat{c}_{k + 1}^\dagger$ and $\hat{ \bar \beta}_B = u\hat{c}_{k + 1} + v\hat{c}_{k}^\dagger$, its orthogonal compliment. Here, $\hat{\bar \beta}_{A}$ corresponds to the jump operator $\betaA$ in the even subspace. 

Next, observe that
\begin{align}
& \Pev \H_S \Pev - (1 - \Pev)\H_S (1 - \Pev) = -2J_k \hat{c}_k^\dagger \hat{c}_{k+1} + \hc  = -2J_k \hat{\beta}_A^\dagger \hat{\beta}_B + \hc 
, 
\end{align}
where $(1 - \Pev)$ projects into the odd parity space. Thus,  while $\Pev \H_S \Pev$ conserves Bogoliubov excitations,  $ (1 - \Pev)\H_S (1 - \Pev)$ may not. This is because $\hat{\bar{\beta}}_B$ is a sum of both $\hat \beta_\gamma$ and $\hat \beta_\gamma^\dagger$. 


\begin{figure}[t]
    \centering
    \includegraphics[width = 3in]{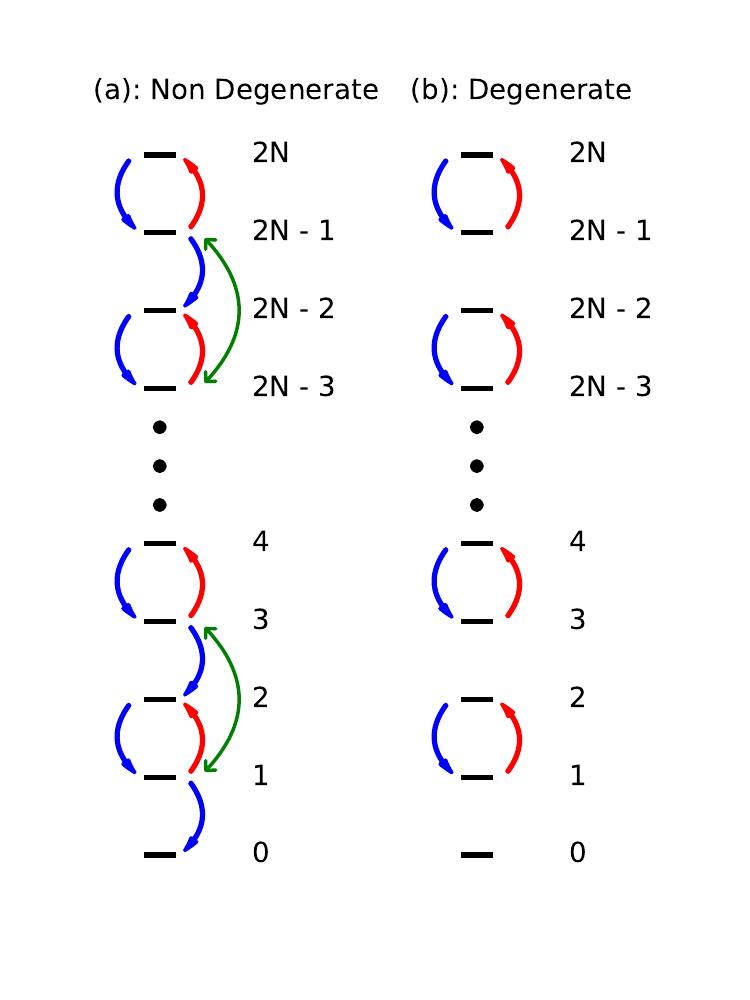}
    \caption{Plotted are the set of $2N + 1$ possible values for the number of Bogoliubov modes excited, where the $m^{\text{th}}$ state level has degeneracy $\binom{2N}{m}$. (a) shows for the non degenerate case where the steady state is unique and corresponds to the vacuum state at the bottom of the tower. The blue arrows show cooling, the red heating, and the green are Hamiltonian interactions. The arrows on the left of the tower begin in even parity states, and the ones on the right side begin in odd parity states. (b) shows the same, except now $u = v$ with a symmetric lattice. This corresponds to odd parity states having only heating and no Hamiltonian interactions, creating $N + 1$ steady states that exist in the manifold spanned by $\{2m,2m-1\}$ Bogoliubov excitations.}
    \label{fig:BogTower}
\end{figure} 


Now, $\hat{\bar \beta}_{A}$ is the jump operators in the even parity sector. In the odd parity sector, the jump operators is
\begin{align}
-\hat{\overline{\beta}}'_{A} &= u\hat{c}_{k} + v\hat{c}_{k+1}^\dagger  = (u^2 - v^2) \hat{\bar \beta}_A + 2uv \hat{\bar \beta}_B^\dagger
, 
\end{align}
and so we see that the jump operator in the odd-parity section is a linear combination of both heating and cooling. When $u \neq v$, the odd parity section has both heating and cooling elements, while the even parity section is strictly cooling. If we construct a tower of Hilbert spaces labeled by the number of Bogoliubov excitations from $0$ to $2N$, then we can break the dynamics down into even and odd parity sections. When the parity is even, the Hamiltonian conserves the number of excitations, and the dissipation only allows for cooling. In the odd parity section, the Hamiltonian can create and destroy pairs of excitations - connecting the levels $2m + 1 \to 2m - 1, 2m + 3$, and the dissipation allows for both heating and cooling of single excitations, connecting the level $2m + 1 \to 2m, 2m +2$. Thus, we can see that every level in the tower can decay to another level except for the manifold with no Bogoliubov excitations. Hence, the unique steady state is the Bogoliubov vacuum, see \cref{fig:BogTower}a.

\subsection{Steady State Degeneracy}
\label{SIsec:2C}

If the Hamiltonian obeys a mirror symmetry, then it turns out that $\hat{\bar \beta}_B$ can be expressed solely as a sum over $\hat \beta_\gamma$; hence the odd parity Hamiltonian becomes Bogoliubov number conserving. Then, exactly when $u = v$, we have that the odd-parity section is \textit{purely} heating, since in that limit we have $\hat{\overline{\beta}}_A' = -\hat{\beta}_B^\dagger$. Since $\H$ conserves Bogoliubov number, we know that if you start with $2m$ Bogoliubov modes, then the dynamics can take you into the space with $2m-1$ excitations through the purely cooling jump operator. Then, you enter the odd parity section, and so now the dynamics can only take you into the space with $2m$ excitations, through the purely heating jump operator. This creates a set of $N + 1$ isolated manifolds, numerically, we find there are exactly $N + 1$ steady states. Further, we know that each steady state must be an incoherent mixture of having $2m$ and $2m-1$ Bogoliubov excitation, or having exactly zero Bogoliubov excitations. Hence, the only \textit{pure} steady state is the Bogoliubov vacuum. Finally, we can see that the only way to end up in the Bogoliubov vacuum is to start there since $\text{ceil}(N_{\text{Bog}}/2)$ is conserved, where ceil rounds up to the nearest integer, and $N_{\text{Bog}}$ is the number of Bogoliubov excitations, see \cref{fig:BogTower}b. This emergent degeneracy is the cause for the long time scale dynamics observed in the main text.

\subsection{Qubits in 2D}
\label{SIsec:2D}

\begin{figure}[H]
\centering
\begin{subfigure}
    \centering
    \includegraphics[width = 3in]{nonRainbowA.pdf}
\end{subfigure}
\begin{subfigure}
    \centering
    \includegraphics[width = 3in]{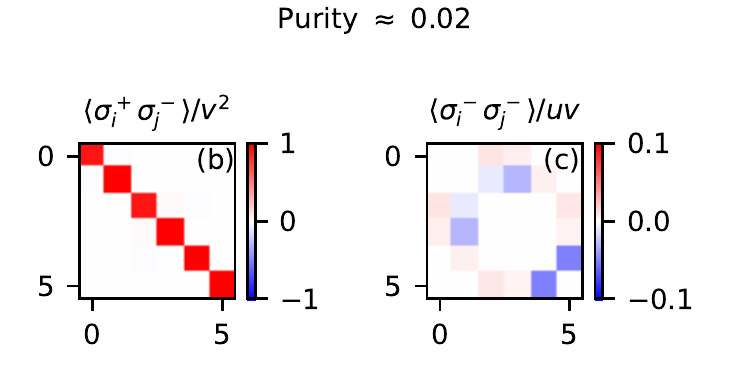}
\end{subfigure}
\caption{\footnotesize{(a) A 2D nearest neighbor XY spin Hamiltonian with two sites coupled to an engineered reservoir. The master equation in identical to that in \cref{fig:nonRainbow}, where we have simply replaced $\hat c_i^{(\dagger)} \to \hat \sigma_i^{\pm}$. (b) and (c) show the spin correlators in the steady state, and also notes that the purity in the steady state $\tr( \rho^2) \approx 0.02$. Despite being the same lattice structure as the fermion in \cref{fig:nonRainbow}, the steady state is no longer pure and has vastly different anomalous correlators.}}
\label{fig:2DQubits}
\end{figure}

It was shown in \cref{SIsec:1C} that the fermions relaxed into pure states regardless of lattice dimension. It is not immediately clear whether or not this would work for qubits, though. To identify the qubit steady state, we relied heavily on using the Jordan-Wigner transformation from the qubits into non-interacting fermions. In a 2D lattice, there is no way for the Jordan-Wigner string operators to drop out of the fermion Hamiltonian, meaning the qubits will always be represented by interacting fermions. For this reason, one does not get the same steady state when going from 2D fermions to 2D qubits. A very simple way to see this is shown in \cref{fig:2DQubits}, where we repeat the exact same structure shown in \cref{fig:nonRainbow}, but use spin operators in place of fermionic ones. The steady state is no longer pure, and the anomalous correlators of the qubits $\langle \hat \sigma_i^- \hat \sigma_j^- \rangle$ are very different than those of the fermions $\langle \hat c_i \hat c_j \rangle$.


\section{Experimental Implementation}
\label{SIsec:3}

\subsection{Sideband Driving}
\label{SIsec:3A}

The correlated dissipator required in our scheme can be readily generated by coupling the two qubits to a single lossy cavity mode (i.e.~an engineered reservoir), and driving the appropriate sideband processes. There are various possible ways to implement the sideband driving: one could either drive the coupling strengths between the qubits and the cavity, or modulate the qubit frequency. Here we present a protocol that makes use of qubit frequency modulation, which can be realized via e.g.~flux tunable transmon qubits.

\begin{figure}[t!]
\centering
\includegraphics[width = 6in]{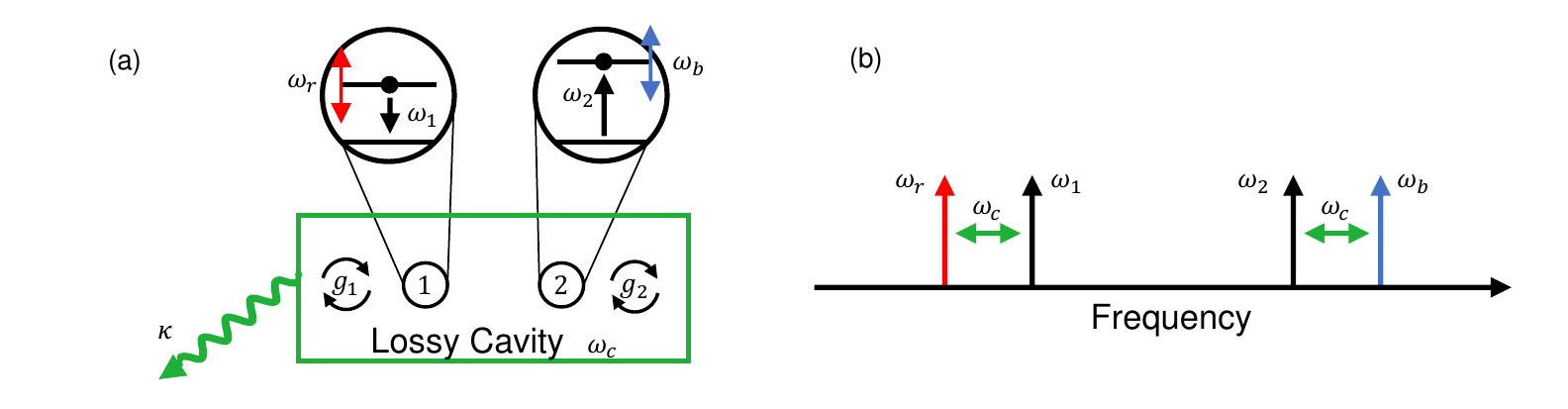}
\caption{\footnotesize{
(a) By placing two qubits $1,2$ with frequencies $\omega_{1,2}$ in a lossy cavity with frequency $\omega_c$ and decay rate $\kappa$, then modulating the qubit frequencies with red and blue sidebands $\omega_{r,b}$, we can engineer the desired dissipation. (b) Shows a schematic of the necessary frequencies needed for the design. 
}}
\label{fig:experiment}
\end{figure}

We consider two qubits (Pauli operators $\hat \sigma_1^z $ and $ \hat \sigma_2^z $) coupled to a lossy cavity mode (annihilation operator $\hat{a}$, frequency $\omega_c$, loss rate $\kappa$). The qubit 1 (2) has resonance frequency $\omega_{1}$ ($\omega_{2}$) and is coupled to the cavity mode with coupling strength $g_{1}$ ($g_{2}$), and we drive the frequencies of qubits 1 and 2 at the red and blue sideband frequencies $\omega_r = \omega_1 - \omega_c$ and $\omega_b = \omega_2 + \omega_c$, respectively (\cref{fig:experiment}). The system Hamiltonian in the lab frame is given by   
\begin{align}
    \H &= \omega_c \hat a^\dagger \hat a + \frac{1}{2}(\omega_1 + \xi_1 \omega_r\cos(\omega_r t)) \hat \sigma_1^z  + \frac{1}{2}(\omega_2 + \xi_2 \omega_b\cos(\omega_b t)) \hat \sigma_2^z  +  (\hat a  + \hat a^\dagger)(g_1 \hat \sigma^x_1 + g_2 \hat \sigma^x_2) 
    . 
\end{align}
$\xi_{1,2}$ are dimensionless drive strengths. Going into the rotating frame defined by the unitary transformation
\begin{align}
    \U &= \exp \left\{ -i \omega_c t \hat a^\dagger \hat a - \frac{i}{2}
    [\omega_1 t + \xi_1 \sin(\omega_r t)] 
    \hat \sigma_1^z  - \frac{i}{2}
    [\omega_2 + \xi_2 \sin(\omega_b t)] \hat \sigma_2^z  \right\}
    , 
\end{align}
the rotating frame Hamiltonian $  \H' $ is given by  
\begin{align}
    \H' &= \U^\dagger \H \U - i\U^\dagger \dot{\U} \\
    &= g_1\left(e^{-i\omega_c t}\hat a + e^{i\omega_ct} \hat a^\dagger \right) 
    e ^{ -i\omega_1 t 
    - i \xi_1 \sin(\omega_r t) } \hat \sigma_1^- 
    + g_2\left(e^{-i\omega_c t}\hat a + e^{i\omega_ct} \hat a^\dagger \right)
    e ^{ -i\omega_2 t 
    - i \xi_2 \sin(\omega_b t) } \hat \sigma_2^-
     + \hc
\end{align}
Making use of the identity 
\begin{align}
    e^{i \xi \sin(\omega t)} &= \sum_{n = -\infty}^\infty J_n(\xi) e^{i n \omega t}
    , 
\end{align}
where $J_n (\xi)$ are Bessel functions, we can rewrite the rotating frame Hamiltonian as 
\begin{align}
    \H' &= 
    \left(e^{-i\omega_c t}\hat a +
    e^{i\omega_ct} \hat a^\dagger \right) 
    \sum_{n = -\infty}^{\infty} 
   \left[ g_1
   J_n(\xi_1)
   e^{ - i(  \omega_1 +  n\omega_r )t } \hat \sigma_1^- + g_2
   J_n(\xi_2)
   e^{ - i(  \omega_2 +  n\omega_b )t } \hat \sigma_2^-  
   \right] + \hc 
\end{align}
Since we drive the qubits at sideband resonances $\omega_r = \omega_1 - \omega_c$ and $\omega_b = \omega_2 + \omega_c$, the rotating terms are resonant when $n= -1$. We can rewrite this as
\begin{align}
    \H' &= \hat{a}^\dagger(g_1' \hat \sigma_{1}^- + g_2' \hat \sigma_2^+) + \hc + \hat V(t)
    , \label{eqn:RWA}
\end{align}
where $g_{ i }' = J_{-1}(\xi_{i })g_{i }$ ($i=1,2$), and $\hat V(t)$ contains only rapidly oscillating terms at frequencies $\omega \gg g_1', g_2'$. Therefore, we can make the standard rotating wave approximation, and treat $\hat V(t)$ as a perturbation of order the coupling strength divided by the oscillation frequency, and neglect it \cite{Jaynes1963, Shore1993}. This approximation has also been shown to match experiment extremely well \cite{Rempe1987}. 

Now that we have found the appropriate Hamiltonian for our system, we can consider the dissipative dynamics. We have considered the cavity mode $\hat a$ is leaky with a loss rate $\kappa$, which can be expressed using a master equation of the form of \cref{eqn:master}:

\begin{align}
    \dot{\hat{\rho}} &= -i[\H',\hat \rho] + \kappa \D[\hat a] \hat \rho.
\end{align}

From here, we trace out the cavity, leaving an effective master equation for only the spins. To do this we will assume that $\kappa \gg g_1',g_2'$, meaning that the cavity is highly damped. Using standard elimination techniques \cite{Gardiner2004,Gardiner1985}, we are then left with the effective dynamics for $\hat \chi = \tr_{\text{cav}} \hat \rho$:

\begin{align}
    \dot{\hat{\chi}} &= \frac{4g^2}{\kappa}\D[\hat L] \hat \chi, \\
    \hat L &= u \hat \sigma_1^- + v\hat \sigma_2^+,
\end{align}

where $g^2 = g_1'^2 + g_2'^2$ and $u = g_1'/g, v = g_2'/g$. This generalizes the result in \cite{SchusterPRL2017}, and agrees exactly in the case $v^2 = 0$. Intuitively, the effective dynamics tell us that the only resonant interactions where the cavity gains a photon is either when spin 1 loses a photon or spin 2 gains a photon. Therefore, if we assume the cavity is highly damped and almost always in the ground state, whenever it loses a photon, it means that one of those two resonant processes occurred, but we have lost the ``which spin'' information. This gives the coherent sum of the two operators. The rate $\frac{4g^2}{\kappa}$ can be derived from a simple Fermi's golden rule calculation, as is shown in \cite{SchusterPRL2017}.


\subsection{Additional Dissipation}
\label{SIsec:3B}

\begin{figure}[t]
\centering
\includegraphics[width = 3.25in]{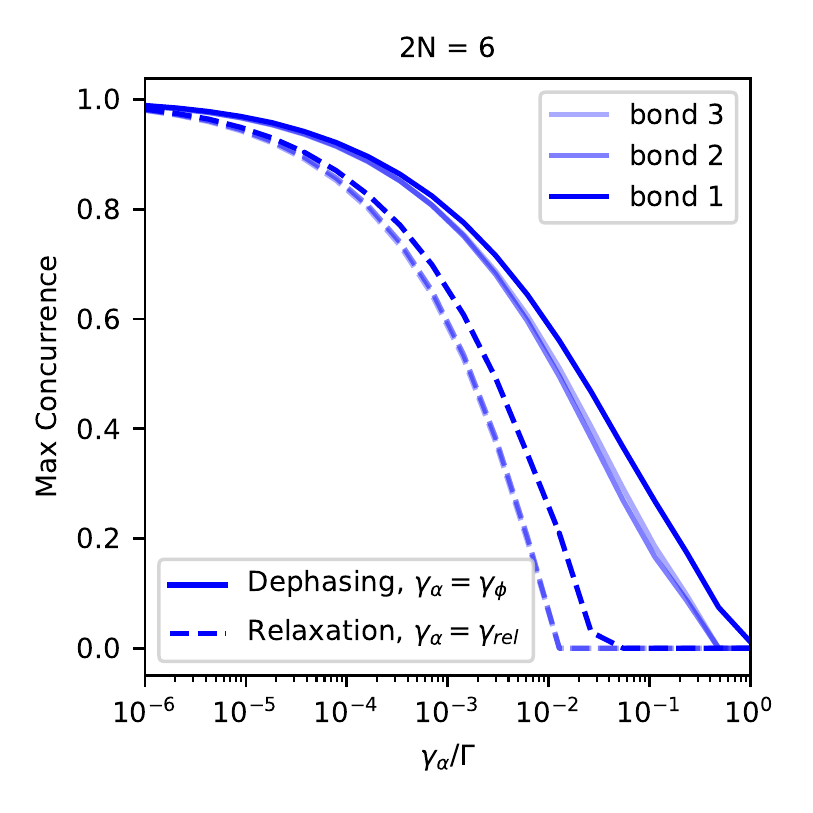}
\includegraphics[width = 3.25in]{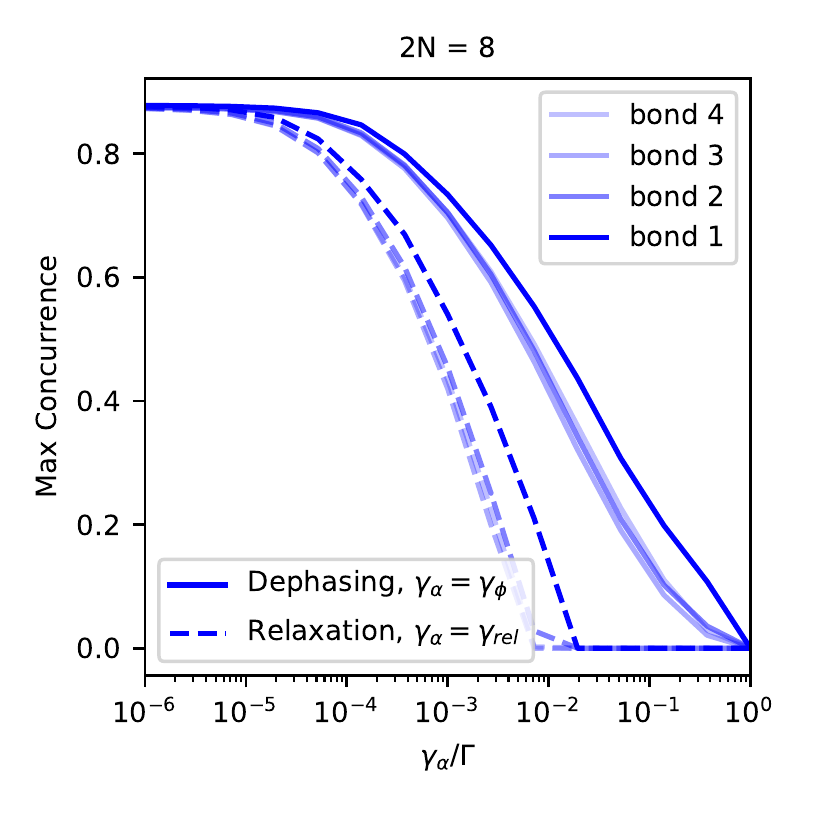}
\caption{Shown is the concurrence across the `rainbows' of the rainbow state with additional dephasing and relaxation for a 6 (left) and 8 (right) site lattice. Note that the first bond, coupled directly to the engineered dissipator, is more persistent, and the other two are identical, despite being different distances from the reservoir. }
\label{fig:AddedDiss}
\end{figure}

We now consider the robustness of our steady state in the presence of additional local dissipation. In \cref{fig:AddedDiss}, we consider the effects of local relaxation (dephasing) with a strength $\gamma_{\text{rel}}$ ($\gamma_\phi$) by plotting the concurrence between pairs of sites $(l,-l)$. In all of the plots, the engineered dissipation couples to the first bond with a strength $\Gamma$, and the Hamiltonian has hopping strength $J = \Gamma$. The master equation is now given by 
\begin{align}
    \L[\hat \rho] &= -i[\H,\hat \rho] + \Gamma\left( \D[\hat \beta_A] +  \D[\hat \beta_B] \right) \hat \rho + \sum_i \D[\hat L_i^\alpha]\hat \rho 
    , 
\end{align}
where the qubit lattice Hamiltonian $ \H $ and the dissipator $ \betaA $ is again given by
\begin{align}
\H &= 
    J \sum_j \hat \sigma_j^+ \hat \sigma_{j + 1}^-  + \hc 
    , \quad
\betaA  = u\hat{\sigma}_{-1}^- + v\hat \sigma_1^+ 
    .
\end{align}
The local dephasing (relaxation) dissipator $\hat L_i^\alpha$ for $ \alpha = \phi$ ($ \alpha = \text{rel} $) are given by
\begin{align}
\hat L_i^{\phi} &= \sqrt{\gamma_\phi} \hat \sigma^z_i , \quad 
\hat L_i^{\text{rel}} = \sqrt{\gamma_{\text{rel}}} \hat \sigma^-_i
    .
\end{align}
We choose $u,v$ to maximize concurrence on each of the entanglement bonds, starting from the 1st, innermost bond, to the $N$th, outermost bond in a $2N$ particle lattice.  

Taking data from a recent publication characterizing a flux-tunable transmon qubit array \cite{Ma2019}, we can estimate the capability of current experiments to perform our protocol. We will assume the transmon qubits are flux-tunable from $2 \pi \times 3.5-6$ GHz, with $\gamma_{\text{rel}}/2\pi  \sim 4-8 $kHz and $\gamma_\phi/2 \pi \sim 40-80$kHz, and tunnel coupling of up to $J/2 \pi \sim 10$MHz. We will take the lossy cavity to couple to the qubits with strength of $g/2 \pi = 15-20$ MHz, and have a decay rate $\kappa/2 \pi \sim 10$ MHz.

With these parameters, assuming a qubit frequency of 4.75 GHz and a cavity frequency of 2.75 GHz, we get red and blue sideband frequencies of $\omega_{r(b)} = 2 (7.5)$ GHz. Given the tunable range of the qubits, the maximum amplitude of the drives is 1.25GHz, giving $\zeta_{1(2)} \approx 0.63 (0.17)$. This gives the maximal values of $g'_{1(2)} \approx 5.95 (1.66)$MHz. This would give an upper bound of $g_{\text{eff}} \approx 6$ MHz. It would therefore be very reasonable to have $\Gamma_{\text{eff}}$ on the order of a few MHz, making $\gamma_{\phi}/\Gamma_{\text{eff}} \sim O(10^{-2})$ and $\gamma_{\text{rel}}/\Gamma_{\text{eff}} \sim O(10^{-3})$. From \cref{fig:AddedDiss}, we can see that this would put an 8 qubit chain within reach of current experiments.

\bibliography{ref}